# COVID-19 Induced Economic Uncertainty: A Comparison between the United Kingdom and the United States

**Ugur Korkut Pata**[1]

Faculty of Economics and Administrative Sciences, Department of Economics, Osmaniye Korkut Ata University, 80000 Merkez/Osmaniye

**Abstract**

The purpose of this study is to investigate the effects of the COVID-19 pandemic on economic policy uncertainty in the US and the UK. The impact of the increase in COVID-19 cases and deaths in the country, and the increase in the number of cases and deaths outside the country may vary. To examine this, the study employs bootstrap ARDL cointegration approach from March 8, 2020 to May 24, 2020. According to the bootstrap ARDL results, a long-run equilibrium relationship is confirmed for five out of the 10 models. The long-term coefficients obtained from the ARDL models suggest that an increase in COVID-19 cases and deaths outside of the UK and the US has a significant effect on economic policy uncertainty. The US is more affected by the increase in the number of COVID-19 cases. The UK, on the other hand, is more negatively affected by the increase in the number of COVID-19 deaths outside the country than the increase in the number of cases. Moreover, another important finding from the study demonstrates that COVID-19 is a factor of great uncertainty for both countries in the short-term.

[1] Corresponding author. E-mail address: korkutpata@ktu.edu.tr; korkutpata@osmaniye.edu.tr

## Introduction

The outbreak of the coronavirus (COVID-19) pandemic has become the focal point of the world. Although the ultimate impact of the pandemic is not yet fully known, it is a problem beyond important issues such as wars, natural disasters, environmental pollution and absence. The COVID-19 virus emerged in Wuhan, China towards the end of 2019, spreading rapidly from person to person and surrounded the world. By May 25, 2020, although China seems to have overcome COVID-19, the number of cases and deaths continues to increase in countries such as the US, Brazil, the UK and Spain. The Chinese government has managed to prevent the spread of the pandemic by implementing quarantine in Wuhan, and temporarily stopping production activities in most sectors. Similarly, many countries have started to take measures with the same practices. However, the lockdowns and quarantines are not economically sustainable and for this reason the measures have started to be stretched gradually.

The COVID-19 is a multi-faceted global crisis that simultaneously interrupts supply, demand and productivity. It has launched a de-globalization process, forcing countries to close their borders, resulting in the restriction of the flow of capital, goods and services between countries, and shot down of business and production, albeit temporarily (Barua, 2020). As the number of COVID-19 deaths and cases increases, uncertainty, panic, fear and anxiety continue to spread in countries. A major uncertainty is associated with COVID-19 cases and deaths. In addition, the global economic practices and related policy reactions of the pandemic are also uncertain (Barro et al. 2020). While some countries can effectively treat reported cases, it is uncertain where, when and how new cases will occur (Mckibbin and Fernando, 2020), how long COVID-19 will last, whether there will be vaccine against the pandemic, how long will governments maintain current incentives to address COVID-19 induced economic problems, and what the social, economic and political implications of this virus will be in the future. It remains unclear whether the economic recovery or stagnation process will be L-, V or U-shaped in many countries after the virus. The pandemic has triggered many more health, social and economic uncertainty. COVID-19 has a negative impact on many sectors including finance, banking, travel, health, service, transportation and infrastructure. The continuity of production and consumption activities in these sectors has been deteriorated due to the virus. Investors stop their investments and withdraw new investment decisions in countries where the virus creates economic uncertainty. In order to protect the citizens of the countries, the social and economic activities against the countries where the virus spreads rapidly are suspended, albeit for a short time. Although the COVID-19 started in China, as of May 2020, the US became the epicenter of the virus. These two countries make up almost 40% of the world's gross domestic product. For the past fifty years, China and the US have not faced a combined supply and demand shock simultaneously. The potential effects of the COVID-19 crisis are much larger than any other seen in history (Fernandes, 2020). For all these reasons, the world is in a period of great uncertainty that has never been seen before.

This study attempts to quantify the impact of COVID-19 on economic policy uncertainty (EPU) in the UK and the US. These two countries are selected due to the availability of daily EPU data. The study focuses on two research questions. I) Does the COVID-19 pandemic affect economic policy uncertainty more negatively in the US or the UK? II) Is the number of cases and deaths in the country or outside the country affecting the EPU more? The answer to these two questions has been sought through empirical analysis.

The rest of the study is structured as follows. Section 2 provides basic information on COVID-19 and economic uncertainty in the UK and the US. Section 3 introduces the data set, models and methods used in the study. Section 4 summarizes and discusses the empirical findings, and the last section concludes the study.

**The COVID-19 and Economic Policy Uncertainty in the US and the UK**

The COVID-19 cases and deaths continue to increase around the world. On May 25, the number of COVID-19 cases exceeded 5,3 million. At the same time, 342,894 people lost their lives due to the pandemic. The COVID-19 mortality rate is approximately 6.4% worldwide. In other words, 6 out of 100 cases die from the virus. The number of COVID-19 deaths and cases in the US is 97,720 and 1,6 million people, respectively. The US accounts for 28% of the worldwide deaths and 30% of the cases caused by COVID-19. Although the US has the highest COVID-19 cases and deaths in the world, considering the mortality rates, the situation is not that bad. On May 25, the mortality rate from COVID-19 was 6% in the US, while it was 14% in the UK. On the same date, the number of COVID-19 cases and deaths in the UK were 259,559 and 36,793, respectively. In terms of infected patients, the virus was transmitted to 0.49% of the US population and 0.38% of the UK population. Neither country has been able to produce any permanent solutions to reduce the spread of the pandemic. In the UK, Prime Minister Boris Johnson was infected with the virus and remained in intensive care for three days. Horton (2020) stated that the UK is following an astonishingly haphazard approach to managing the COVID-19 crisis. There are similar problems in the US. The International Monetary Fund (2020) foresees that the economies of the US and the UK will contract by 5.9% and 6.5% in 2020. Governments' failure to tackle COVID-19 continues to adversely affect economies. According to Baker et al. (2020a), a year-on-year contraction in the US economy will be around 11%-20% in the last quarter of 2020. The authors stated that half of this contraction in the output of the US would be due to COVID-19 induced uncertainty. Moreover, Baker et al (2020b) predicted that the GDP growth of the US will decrease by 7% in the second quarter of 2020. Similarly, Dietrich et al. (2020) stated that US households expect a 6% decrease in output within the 12 months following March 2020 and also documented that the uncertainty in output loss is quite large. Ludvigson et al. (2020) also argued that the US industrial production will decrease by 12.75% and service sector employment by 17% over a period of ten months, and the COVID-19 induced macroeconomic uncertainty will last for five months. As stated in the above studies, in order to eliminate the negative effects of increasing uncertainty on the economy, the FED tried to revive the falling aggregate demand by cutting interest rates.

In addition to the China-US trade wars, the Brexit process and the conflicts in the middle east, uncertainty spiked as a result of the rapid spread of the COVID-19 virus (Leduc and Liu, 2020). The COVID-19 pandemic caused a large increase in uncertainty, similar to the 1929 Great Depression rather than the 2008 global crisis (Baker et al. (2020a). Empirically, uncertainty causes significant declines in production, consumption and investment activities, and the peak of this negative situation appears exactly one year later (Basu and Bundick, 2017). In the COVID-19 era, economic policy uncertainty has increased significantly in the UK and the US. This is shown in Fig. 1.

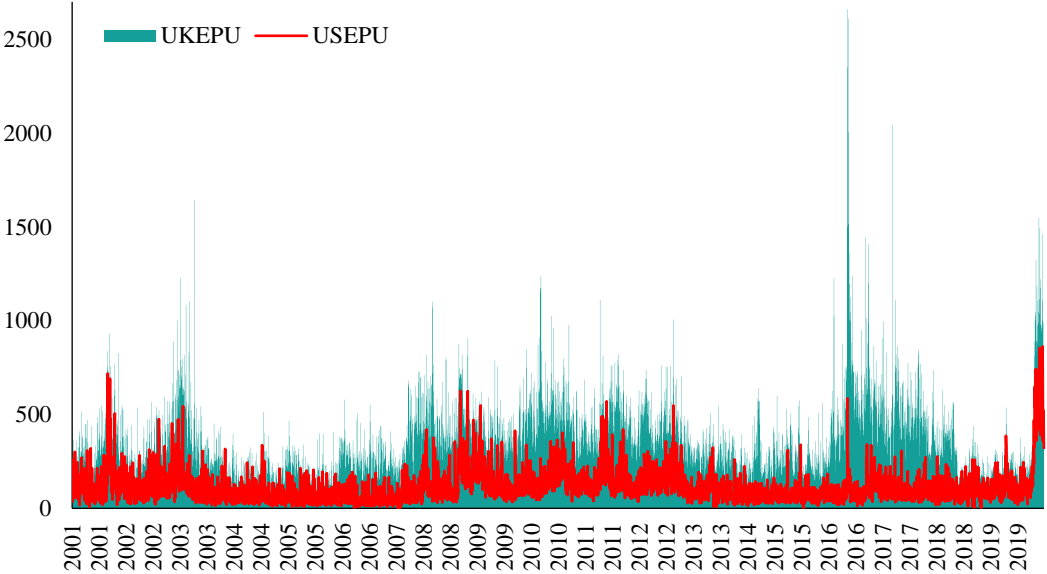

**Fig. 1** EPU in the US and the UK from 1 January 2001 to 25 May 2020.

Fig. 1. illustrates the time path of EPU in the US and the UK over a 20-year period. The US has been experiencing the highest economic uncertainty of the last 20 years. In 2003, when the US and the UK invaded Iraq and then the Iraq disarmament crisis occurred, the economic uncertainty of the UK is high. In 2016, when the UK EPU reached the highest point, the British

government held a referendum to exit the European Union. The UK officially left the European Union on January 31, 2020. The Brexit process is expected to fully end on December 31, 2020. Recently, besides Brexit, the COVID-19 pandemic has also caused an increase in EPU. Since the virus appeared, the value of EPU in the UK has risen 15 times to over 1,000.

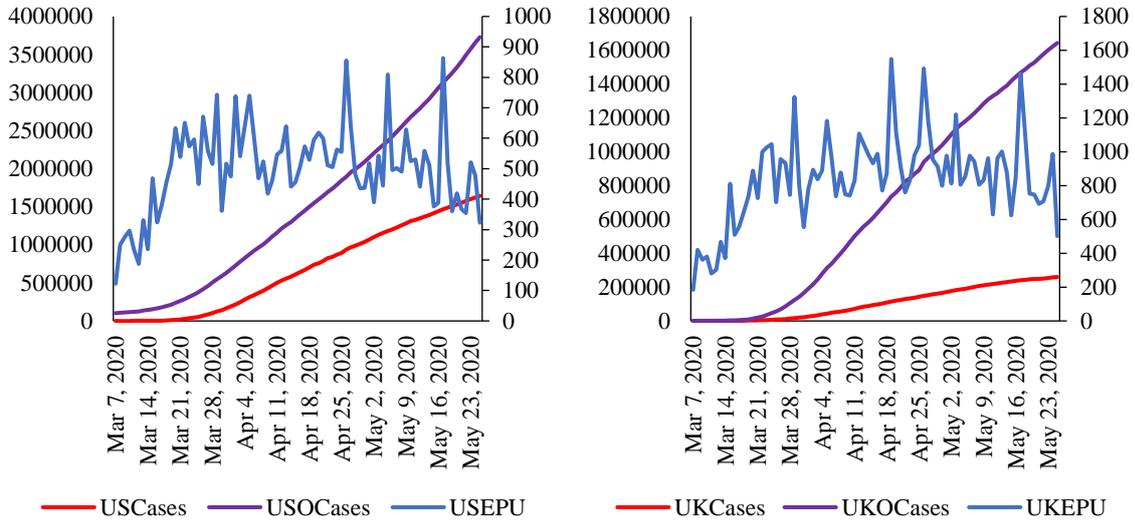

**Fig. 2** EPU, COVID-19 cases in the US, the UK and outside the countries.

Fig. 2 shows the EPU and COVID-19 cases in the US and the UK, and COVID-19 cases outside of the countries. In the figure, the left vertical axes indicate the COVID-19 statistics, and the right vertical axes indicate the EPU statistics. Accordingly, the EPU values, which were around 200 in the first week of March in both countries, increased to 700 and 1300 in the US and UK towards the end of March, respectively. Moreover, EPU reached high levels in both countries in April and May.

**Data Set, Model and Methodology**

In this study, we investigated the effect of COVID-19 on EPU in the US and the UK covering the period of March 7, 2020 to May 24, 2020. To this end, we used four different models as follows:

$$\ln EPU_{US_t} = \beta_0 + \beta_1 \ln cases_{US_t} + u_t \quad (\ln EPU_{UK_t} = \beta_0 + \beta_1 \ln cases_{UK_t} + u_t) \quad (1)$$

$$\ln EPU_{US_t} = \delta_0 + \delta_1 \ln cases_{OUS_t} + e_t \quad (\ln EPU_{UK_t} = \delta_0 + \delta_1 \ln cases_{OUK_t} + e_t) \quad (2)$$

$$\ln\text{EPU}_{US_t}=\mu_0+\mu_1\ln\text{deaths}_{US_t}+v_t \quad (\ln\text{EPU}_{UK_t}=\mu_0+\mu_1\ln\text{deaths}_{UK_t}+v_t) \qquad (3)$$

$$\ln\text{EPU}_{US_t}=\tau_0+\tau_1\ln\text{deaths}_{OUS_t}+z_t \quad (\ln\text{EPU}_{UK_t}=\tau_0+\tau_1\ln\text{deaths}_{OUK_t}+z_t) \qquad (4)$$

where $\beta_0$, $\delta_0$, $\mu_0$, and $\tau_0$ are the constant terms, $\beta_1$, $\delta_1$, $\mu_1$, and $\tau_1$ are the long-term coefficients, $u_t$, $e_t$, $v_t$, and $z_t$ are the independent and identically distributed error terms, $\text{EPU}_{US}$ ($\text{EPU}_{UK}$) refers to daily economic policy uncertainty in the US (the UK), $\text{cases}_{US}$ ($\text{cases}_{UK}$) represents the number of COVID-19 cases in the US (the UK), $\text{deaths}_{US}$ ($\text{deaths}_{UK}$) denotes the number of COVID-19 deaths in the US (the UK), $\text{cases}_{OUS}$ ($\text{cases}_{OUK}$) illustrates the number of COVID-19 cases outside the relevant country, and $\text{deaths}_{OUS}$ ($\text{deaths}_{OUK}$) indicates the number of COVID-19 deaths outside the relevant country. The EPU index is calculated with the frequency of use of the terms about economics, policy and uncertainty in daily newspaper articles. The data of COVID-19 deaths and cases were obtained from European Centre for Disease Prevention and Control (2020), and the EPU data were collected from https://www.policyuncertainty.com/index.html developed by Baker et al. (2016). Before performing the analysis, all variables were converted into logarithmic form.

We used the bootstrap autoregressive distributed lag (ARDL) approach in order to examine short- and long-term relationships between variables. In this approach, in addition to the conventional bounds testing procedure, a new test developed by McNown et al. (2018) and critical values are obtained with the bootstrap resampling method. For the bootstrap ARDL approach, in the first step, the following unrestricted error correction model (UECM) proposed by Pesaran et al. (2001) can be estimated as follows:

$$\Delta ln\text{EPU}_t = \beta_0 + \sum_{i=1}^{n}\alpha_1 \Delta ln\text{EPU}_{t-i} + \sum_{i=0}^{m}\alpha_2 \Delta ln\text{cases}_{t-i} + \mu_1 ln\text{EPU}_{t-1} + \mu_2 ln\text{cases}_{t-1} + u_t \qquad (5)$$

where $\Delta$ is the difference operator, $\beta_0$ is the constant term, n and m are indices of lags, $ln\text{EPU}_t$ is the dependent variable, $ln\text{cases}_t$ is the independent variable, $\mu_1$ and $\mu_2$ are the long term coefficients, $\alpha_1$ and $\alpha_2$ are the short term coefficients, and $u_t$ is the i.i.d. error term. This equation represents the first stage of the conventional ARDL bounds test that widely used by researchers because it allows analysis of variables that have different order of integration. In this approach, Pesaran et al. (2001) proposed two tests for the analysis of cointegration. From these tests, the overall F-test is applied to both independent and depended variables, while the t-test (t-dependent) is applied only to the dependent variable.

Pesaran et al. (2001) stated that degenerate cases occur in two ways depending on whether the dependent variable or independent variables are found to be statistically insignificant in the error correction model (Degenerate case I: The test statistics of the overall F-test and t-dependent test are significant, while the independent variable is stationary at level. Degenerate case II: The overall F-test is significant but t-dependent test is insignificant). In the degenerate cases, there is no cointegration between the variables in the model. On the other hand, in the conventional ARDL approach, there are two conditions for the identification of cointegration. First, the null hypothesis of no cointegration must be rejected with the overall F- and the t-dependent tests. As the second condition, the dependent variable must be I (1). Pesaran et al. (2001) excluded the degenerate case I, assuming the dependent variable I (1). However, this exclusion and a test based only on the overall F statistic can cause biased results. Therefore, McNown et al. (2018) proposed an additional test applied only to the level values of the independent variables. Thus, the significance of the dependent variable and independent variables in the error correction term can be tested separately. Moreover, the authors derived critical values using the bootstrap method. The bootstrap test improves the estimation in terms of size and power properties. Furthermore, the bootstrap method maintains its strong size and properties even when all variables are I(0) (McNown et al., 2018). This method also eliminates the inconclusive inferences about the existence of cointegration and degenerate cases (Goh et al. 2020), and the requirement of the dependent variable to be I(1). The bootstrap ARDL approach is unique because it is not sensitive to the integration properties of the variables (Nawaz et al. 2019). Furthermore, this approach provides bootstrapped critical values for overall F, t-dependent and F-independent tests. The null and alternative hypotheses for three test statistics can be written as follows:

- For overall F-test statistic, $H_0: \mu_1=\mu_2=0$, $H_{alternative}: \mu_1 \neq \mu_2 \neq 0$
- For t-dependent test statistic, $H_0: \mu_1=0$, $H_{alternative}: \mu_1 \neq 0$
- For F-independent test statistic, $H_0: \mu_2=0$, $H_{alternative}: \mu_2 \neq 0$

The null hypotheses are rejected if the test statistics of the overall F and F-independent tests are greater than the bootstrap critical values. For the t-dependent test, the null hypothesis is rejected when the test statistic is less than the relevant critical value. If the null hypothesis is rejected with all three test statistics, then the exact cointegration relationship is confirmed. After confirming cointegration, short- and long-term coefficients are estimated simultaneously.

**Empirical Results and Discussion**

In the first stage of the analysis, we investigated the stationarity properties of the variables to ensure that none of the variables are integrated at I(2). The bootstrap ARDL test can be used without knowing the stability properties of the series. However, the critical values considered in this approach are derived by the assumption that the variables are stationary at the level or first difference. If any variable is second difference stationary, the bootstrap ARDL approach cannot be applied. In order to determine if this condition is met, we used conventional Augmented Dickey-Fuller (ADF) (Dickey and Fuller, 1981) and Phillips-Perron (PP) (Phillips and Perron, 1988) unit root tests. The results of the unit root tests are given in Table 1.

**Table 1** Unit Root Tests Results

| Test | ADF | | PP | |
|---|---|---|---|---|
| Variables | Level | First difference | Level | First difference |
| $lnEPU_{US}$ | -5.072* | – | -4.921* | – |
| $lnEPU_{UK}$ | -4.279* | – | -4.093* | – |
| $lncases_{US}$ | -11.231* | – | -4.431* | – |
| $lndeaths_{US}$ | -0.755 | -3.843** | -0.643 | -8.537* |
| $lncases_{OUS}$ | -10.598* | – | -4.335* | – |
| $lndeaths_{OUS}$ | -8.123* | – | -0.159 | -3.513** |
| $lncases_{UK}$ | -4.536* | – | -4.398* | – |
| $lndeaths_{UK}$ | -2.670 | -6.054* | -7.110* | – |
| $lncases_{OUK}$ | -11.231* | – | -3.832** | – |
| $lndeaths_{OUK}$ | -6.617* | – | 0.330 | -3.515** |

**Note:** * denotes significance at 1% level.

According to the results shown in Table 1, six of the 10 variables are stationary at the level, while the other four variables contain unit roots. Moreover, none of the variables are stationary at the second difference [I(2)]. Because the variables have a different order of integration and the dependent variables are level stationary, we performed the bootstrap ARDL approach to investigate cointegration between the variables. The results of this approach are shown in Table 2.

**Table 2** Bootstrap ARDL Results

| Models | ARDL (lags) | overall F-stat | 5% CV | t-dep | 5% CV | F-indep | 5% CV |
|---|---|---|---|---|---|---|---|
| 1-) $lnEPU_{US}=f(lncases_{US})$ | 1,1 | 7.353** | 6.004 | -3.767** | -3.117 | 2.495** | 2.480 |
| 2-) $lnEPU_{US}=f(lndeaths_{US})$ | 1,2 | 8.018** | 6.273 | -3.914* | -3.143 | 2.513 | 3.052 |
| 3-) $lnEPU_{US}=f(lncases_{OUS})$ | 1,1 | 7.934* | 5.982 | -3.978* | -3.155 | 2.759** | 2.121 |
| 4-) $lnEPU_{US}=f(lndeaths_{OUS})$ | 1,2 | 10.902* | 6.018 | -4.653* | -3.117 | 3.471** | 2.989 |
| 5-) $lnEPU_{UK}=f(lncases_{UK})$ | 1,1 | 6.365** | 5.634 | -3.524** | -2.900 | 2.127 | 2.429 |
| 6-) $lnEPU_{UK}=f(lndeaths_{UK})$ | 1,1 | 5.593** | 5.555 | -3.086** | -2.780 | 0.982 | 2.083 |
| 7-) $lnEPU_{UK}=f(lncases_{OUK})$ | 1,1 | 8.512* | 5.704 | -4.082* | -3.083 | 2.830** | 2.465 |
| 8-) $lnEPU_{UK}=f(lndeaths_{OUK})$ | 1,2 | 9.689* | 6.813 | -4.387* | -3.262 | 3.437** | 3.380 |

**Note:** The number of bootstrap replications is 2000. * and ** denote significance at 1% and 5% levels, respectively. CV: Critical values.

We used the bootstrap ARDL approach for eight different models. These models were analyzed based on COVID-19 cases and deaths (inside and outside the country). For the bootstrap ARDL models, the optimal lag lengths are selected based on the Schwarz-Bayesian criterion. The results of the bootstrap ARDL test show that there exists a long run relationship between the variables in 5 out of eight models. In the five models, the null hypothesis of no cointegration is rejected by both F-tests and the t-test. Interestingly, in the UK, both the number of cases and deaths within the country do not affect the EPU. In the US, only COVID-19 deaths that occur within the country do not affect EPU. For both countries, COVID-19 cases and deaths outside the relevant country affect its EPU. At the last stage of the analysis, we estimated the long- and short-term coefficients based on the ARDL model. These coefficients are reported in Table 3.

**Table 3** Long- and short-run estimation

| 1-) $lnEPU_{US}=f(lncases_{US})$ | | | |
|---|---|---|---|
| $\Delta lncases_{US_t}$ | $ECT_{t-1}$ | $lncases_{US_t}$ | Constant |
| 2.892* | -0.865* | 0.197* | 3.416* |

| 3-) $lnEPU_{US}=f(lncases_{OUS})$ | | | |
|---|---|---|---|
| $\Delta lncases_{OUS_t}$ | $ECT_{t-1}$ | $lncases_{OUS_t}$ | Constant |
| 7.687* | -0.784* | 0.265* | 1.636* |

| 4-) $lnEPU_{US}=f(lndeaths_{OUS})$ | | | | |
|---|---|---|---|---|
| $\Delta lndeaths_{OUS_t}$ | $\Delta lndeaths_{OUS_{t-1}}$ | $ECT_{t-1}$ | $lndeaths_{OUS}$ | Constant |
| 0.780 | 5.250** | -0.885* | 0.220* | 2.995* |

| 7-) $lnEPU_{UK}=f(lncases_{OUK})$ | | | |
|---|---|---|---|
| $\Delta lncases_{OUK_t}$ | $ECT_{t-1}$ | $lncases_{OUK_t}$ | Constant |
| 2.422* | -0.774* | 0.220* | 2.835* |

| 8-) $lnEPU_{UK}=f(lndeaths_{OUK})$ | | | | |
|---|---|---|---|---|
| $\Delta lndeaths_{OUK_t}$ | $\Delta lndeaths_{OUK_{t-1}}$ | $ECT_{t-1}$ | $lndeaths_{OUK}$ | Constant |
| -1.310 | 6.043* | -0.749* | 0.242* | 2.729* |

**Note:** The number of bootstrap replications is 2000. * and ** denote significance at 1% and 5% levels, respectively. CV: Critical values.

The five models pass all the diagnostic tests for autocorrelation, stability, non-normality, specification and heteroscedasticity (see Table 1A and Fig. 1A in the Appendix). For the US, both domestic and international COVID-19 cases increase the EPU in the long-term. Moreover, the increasing number of COVID-19 deaths in the country raises uncertainty. The uncertainty-enhancing effect of COVID-19 cases outside of the US is greater than the increase in the number of deaths. In the short term, COVID-19 also creates enormous uncertainty for the US economy. When the results are analyzed for the UK, it is seen that the increase in COVID-19 cases and deaths outside of the UK have a positive impact on its EPU. Unlike the US, COVID-19 deaths have a greater impact on increasing EPU in the UK. Keeping other things constant, COVID-19 has a significant effect on EPU for both countries.

To our knowledge, only one study to date has examined the effect of COVID-19 on EPU. Albulescu (2020) investigated the impact of COVID-19 on the EPU in the US covering the period of 21 January 2020 to 13 March 2020, and found that global COVID-19 cases and the death ratio have no effect on the US EPU. However, investigating the situation outside China, he determined that the increase in the number of cases and death ratio have a positive influence on EPU in the US. Differently, we examined the effects of COVID-19 deaths and cases on the EPU in the US and the UK. We also analyzed the effects of total COVID-19 deaths and cases outside the US and the UK on EPU. When Albulescu (2020) conducted the study, there were only 2,126 cases and 48 deaths in the US on March 13. To date, the situation in COVID-19 is quite different in both the US and the UK. The US is the country with the highest number of cases and deaths in the world. The UK ranks 5th in terms of number of COVID-19 cases. Nevertheless, despite differences, our empirical results support the findings of Albulescu (2020) who reported that COVID-19 contributes to uncertainty.

**Conclusion**

The rapid increase in COVID-19 cases and deaths put pressure on financial markets and real economies. In many countries, production and consumption activities in various sectors are decreasing due to stay at home and quarantine measures that are taken to prevent the spread of the pandemic. It remains unclear when the COVID-19 will stop, whether a second wave will be experienced again and what its economic effects will be. This also causes an increase in economic policy uncertainty. To this context, the study has analyzed the effect of COVID-19 on EPU in the US and the UK, comparatively. For that purpose, we perform a bootstrap ARDL approach.

There are three main findings of the study. First, the COVID-19 pandemic increases economic policy uncertainty in both the US and the UK. Second, the short-term adverse effect of the pandemic on uncertainty is much more than the long-term. Third, in terms of economic uncertainty, the UK is more sensitive to COVID-19 deaths than cases. On the contrary, the US is more affected by the increase in the number of COVID-19 cases. The results of the study indicate that the relationship between the pandemic and uncertainty may vary depending on the COVID-19 criterion, the country studied and the situation of COVID-19 in and out of the country.

# Appendix

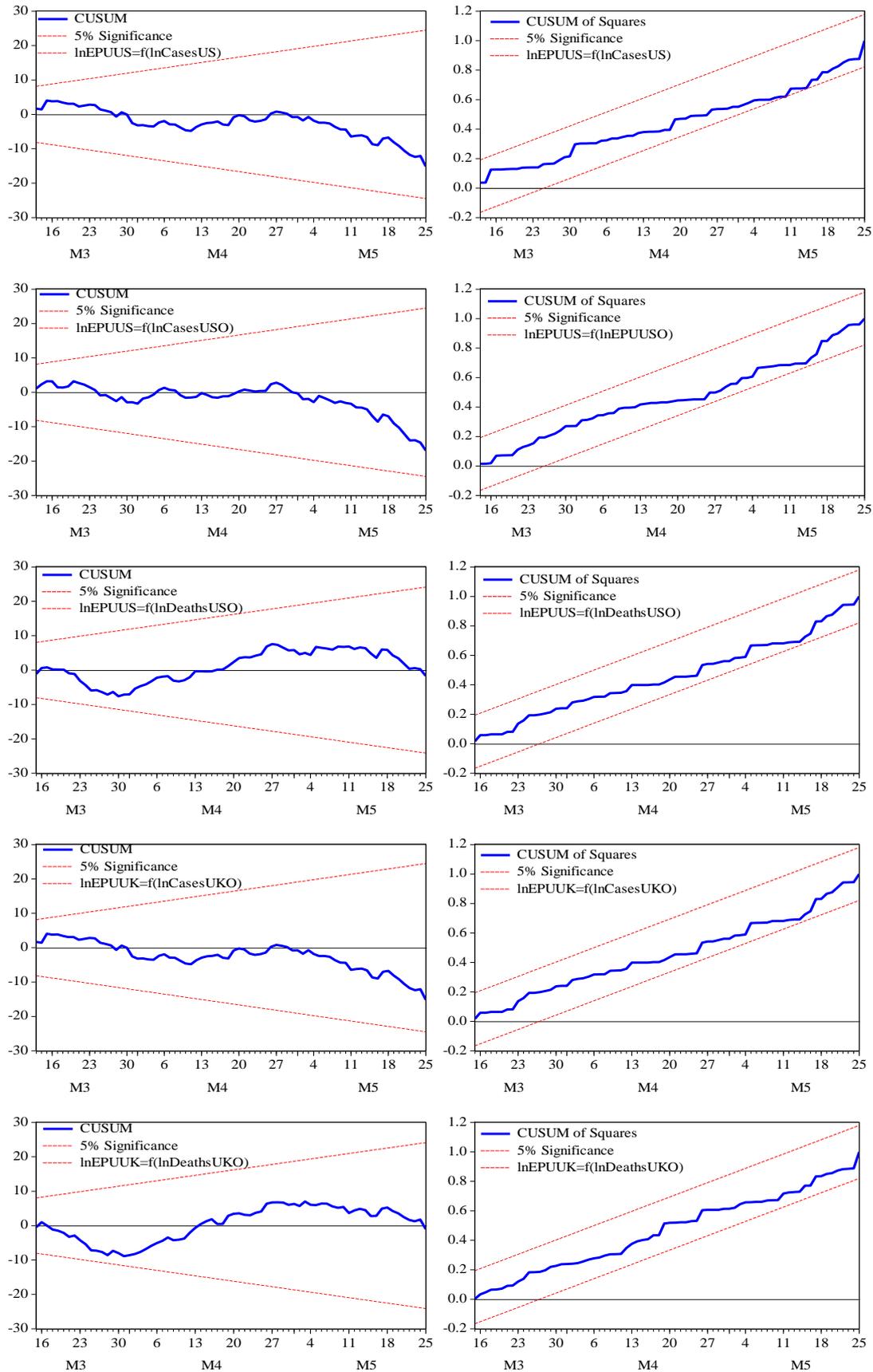

**Fig. 1A** The results of the CUSUM and CUSUMSQ tests

**Table 1A** ARDL diagnostic tests results

| Models | LM | White | Ramsey | JB | ARCH |
|---|---|---|---|---|---|
| 1-) $lnEPU_{US}=$ $f(lncases_{US})$ | 0.037 (0.847) | 2.137 (0.102) | 0.070 (0.791) | 0.171 (0.917) | 0.001 (0.971) |
| 3-) $lnEPU_{US}=$ $f(lncases_{OUS})$ | 2.458 (0.121) | 1.176 (0.324) | 1.584 (0.212) | 0.441 (0.801) | 0.355 (0.552) |
| 4-) $lnEPU_{US}=$ $f(lndeaths_{OUS})$ | 0.669 (0.515) | 1.060 (0.382) | 0.102 (0.750) | 2.565 (0.277) | 0.958 (0.330) |
| 7-) $lnEPU_{UK}=$ $f(lncases_{OUK})$ | 0.115 (0.734) | 0.440 (0.724) | 0.642 (0.425) | 2.883 (0.236) | 1.906 (0.171) |
| 8-) $lnEPU_{UK}=$ $f(lndeaths_{OUK})$ | 0.195 (0.823) | 1.557 (0.195) | 0.213 (0.553) | 0.972 (0.614) | 1.386 (0.242) |

**Note:** ( ): probability values.